# Preliminary corrosion studies of IN-RAFM steel with stagnant Lead Lithium at 550 C

A.Sarada Sree, Hemang S.Agravat, Jignesh Chauhan and E.Rajendrakumar

Institute for Plasma Research, Bhat, Gandhinagar-382428.

**E-mail**: sarada.sree@gmail.com
**Phone :** +917923964032.
**Fax    :** +917923962277

## Abstract

Corrosion of Indian RAFMS (reduced activation ferritic martensitic steel) material with liquid metal, Lead Lithium ( Pb-Li) has been studied under static condition, maintaining Pb-Li at 550 C for different time durations, 2500, 5000 and 9000 hours. Corrosion rate was calculated from weight loss measurements. Microstructure analysis was carried out using SEM and chemical composition by SEM-EDX measurements. Micro Vickers hardness and tensile testing were also carried out. Chromium was found leaching from the near surface regions and surface hardness was found to decrease in all the three cases. Grain boundaries were affected. Some grains got detached from the surface giving rise to pebble like structures in the surface micrographs. There was no significant reduction in the tensile strength, after exposure to liquid metal. This paper discusses the experimental details and the results obtained.

## 1. Introduction

In Indian LLCB ( Lead Lithium ceramic breeder) test blanket module program [1] concept, Pb-Li is used as coolant, neutron multiplier and tritium breeder. Indian RAFMS (reduced activation ferritic martensitic steel) is the candidate structural material for this test blanket module. Compatibility of the structural material with the liquid metal is one of the prime concerns for the successful operation of TBM in ITER [2-4]. Corrosion in the form of dissolution, inter granular penetration and impurity transfer to the liquid metal or from the liquid metal to the structural material, can cause wall thinning, leading to loss of mechanical integrity. Corrosion by the liquid metal could be the limiting factor for the life of the structural material in ITER (international thermo nuclear experimental reactor). Therefore compatibility of the structural material with liquid metal need to be studied in stagnant as well as in dynamic conditions to understand the different factors contributing to corrosion. Corrosion and compatibility of ferritic martensitic steels [5-31] and austenitic steels [32-35] with Pb-Li have been carried out in static and flowing conditions.

Therefore, to study the effect of corrosion of IN-RAFMS in Lead Lithium (Pb-Li) environment, initially a static experiment was planned, in which IN-RAFMS sample coupons were exposed to static liquid metal maintained at 823 K(550 C). Samples were taken out after 2500, 5000 and 9000 hours and analyzed using SEM (Scanning electron microscope). Chemical composition was measured on the cross section of the samples by SEM/EDX (energy dispersive X-ray analysis) measurements. Hardness and tensile testing were also carried out on exposed samples. This paper discusses the experimental details and the results obtained.

## 2. Experimental set up

Experimental set up is shown in figure 1. Lead Lithium chunks are loaded in a cylindrical chamber made of SS316 L. A thermo well was fixed on the side of the chamber. A 'K' type thermocouple was introduced into the thermo well, which extends up to the center of the chamber. A 3 KW heater coil was wound over the circumference of the chamber to melt the Lead-Lithium chunks.

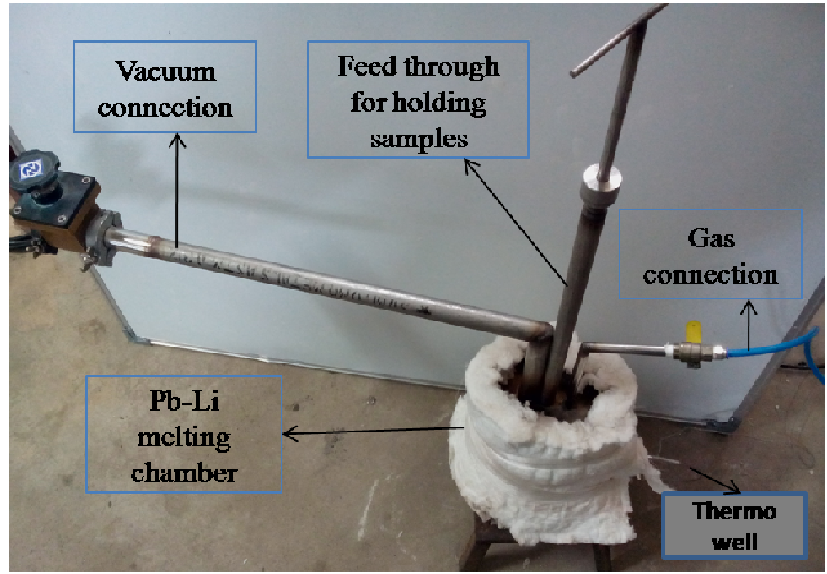

**Fig.1.** Experimental setup for studying corrosion of IN-RAFMS with static Pb-Li at 550 C.

Top flange of the chamber has the provision for the following.
a. to evacuate the chamber
b. to introduce Argon gas into the chamber to maintain positive pressure over liquid metal melt.
c. a movable feed through arrangement to hold the samples.

IN-RAFMS flat and tensile samples were fixed to a sample holder, attached at the end of the feed through. Chemical composition of IN-RAFMS material is given in table 1. Initially samples were held above the Pb-Li chunks. The chamber is evacuated upto $10^{-3}$ m.bar, raising the temperature of the chunks upto 200 C. Then the vacuum system was isolated and Argon gas was introduced into the chamber. Then the temperature was further raised up to 550 C. Once all the chunks were melted and maintained at a temperature of 550 C, the sample holder was lowered with the feed through arrangement to dip the samples in the liquid metal melt. A positive pressure of 1.5 bar is maintained over the liquid melt throughout the experiment.

Samples were removed from the chamber after exposing them to the liquid metal for 2500, 5000 and 9000 hours. To remove the adherent Pb-Li, the samples were cleaned, using cleaning solution consisting of Acetic acid ($CH_3COOH$), Hydrogen peroxide ($H_2O_2$) and Ethyl Alcohol ($C_2H_5OH$) in 1:1:1: ratio. The samples were cleaned and weighed until similar consecutive readings were obtained. Weight measurements were taken using Sartorius precision weighing balance with a precision of ± 0.01 mg. After cleaning, the samples were cut and molds were prepared for metallographic examination. The molds were ground using different grade SiC papers and finally polished with Alumina powder to achieve mirror finishing.

The samples were analyzed using SEM (scanning electron microscope). Change in chemical composition was determined with SEM-EDX (energy dispersive X-ray analysis) measurement. Line scan analysis was carried out on the cross section of the samples. Surface micrographs were obtained using scanning electron microscope, model S440i, from LEO corporation, UK. Micro vickers hardness and tensile strength measurements were also carried out on the samples. Hardness measurements were carried out using a Mitutoyo HM 211, Micro Vickers hardness testing machine. Tensile testing was carried out using Instron make, 5982 Universal testing machine.

**Table 1: Chemical composition of IN- RAFMS (wt%)**

| Cr | C | Mn | V | W | Si | P | S | Ta | Nb | Mo | Ni | Fe |
|---|---|---|---|---|---|---|---|---|---|---|---|---|
| 9.15 | 0.08 | 0.53 | 0.24 | 1.37 | 0.026 | <0.002 | 0.002 | 0.08 | <0.001 | <0.002 | 0.004 | Bal. |

## 3. Results

### 3.1. Weight loss measurements

Weight of the samples was measured before and after exposure to liquid metal at different time intervals. Corrosion rate is calculated from the weight loss measurements and given in table 2.

**Table 2: Corrosion rate after exposure to Pb-Li**

| Sr. No. | Exposure time (hours) | Dissolution rate $g/(m^2 \times year)$ | Corrosion rate (μm/year) |
|---|---|---|---|
| 1 | 2500 | 292.42 | 37.68 |
| 2 | 5000 | 276.85 | 35.67 |
| 3 | 9000 | 184.72 | 23.80 |

From weight loss measurements, corrosion rate of IN-RAFMS was found to be ~ 40 μm/year. Corrosion rate estimated from weight loss measurements was almost found to be same for 2500 and 5000 hours exposures. Decrease in the corrosion rate for 9000 hours exposure could be due to saturation of liquid metal. This could be possible, due to the formation of saturated layer of dissolved elements near the sample surface.

*3.2. SEM surface micrographs*

After exposure to liquid metal, the samples lost their metallic luster and the surface became dull. Surface micrographs of exposed samples for 2500, 5000 and 9000 hours were shown in figure 2(a), (b) and (c) respectively. Surface got deteriorated with increasing exposure time.

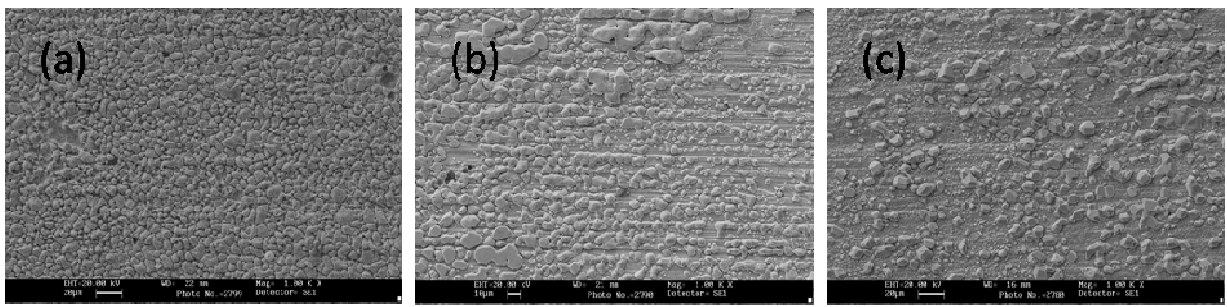

**Fig.2.** SEM surface micrographs of samples exposed to static Pb-Li at 550 C.
Exposure time: (a) 2500 hours, (b) 5000 hours, (c) 9000 hours.

In all the exposed samples, pebble like structure can be seen. Similar type of surface morphology was observed by V.Tsisar et al., [30] when EP-823 steels were exposed to Nitrogen added Lithium maintained at 600 C. This type of structure is due to dislodging of grains/ sub grains from the surface. Initially, when the surface of the structural material is exposed to Lead-Lithium eutectic at 550 C, corrosion occurs on the surface and the bonding between grains / sub grains become weak. Slowly the grains/ sub grains get detached from the surface and go into the eutectic, giving rise to pebble like structure.

*3.3. SEM cross section micrographs*

SEM cross section micrographs after exposure of the samples to static liquid metal at 550 C are shown in figure 3(a), (b) and (c) for the case of 2500, 5000 and 9000 hours respectively.

Groove like structure was observed in all the three cases, showing internal corrosion of the material. After 5000 hours, a detached layer was observed. Chemical composition, measured in the detached layer revealed that it is part of the matrix and contained mainly iron and chromium.

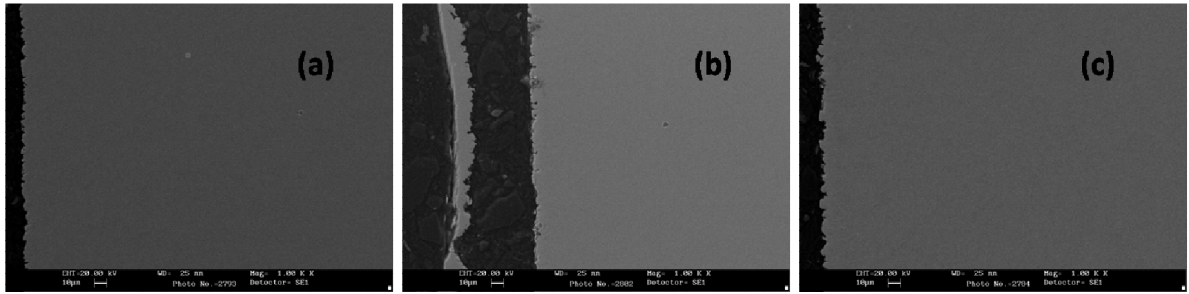

**Fig.3.** SEM cross section micrographs after exposure to static Pb-Li at 550 C.
Exposure time: (a) 2500 hours, (b) 5000 hours and (c) 9000 hours.

*3.4. SEM EDX measurements on cross section*

SEM EDX was carried out on the cross section of the samples exposed for 2500, 5000 and 9000 hours. Chemical composition was measured depth wise and shown in figure 4(a), (b) and (c) for the above mentioned three exposure times. In all the three cases, chromium is found leaching from the near surface regions and tungsten is found slightly high near the surface. In case of 5000 and 9000 hours exposure, chromium is found leaching from a greater depth ~20 $\mu$m as compared to ~5 $\mu$m in case of 2500 hours.

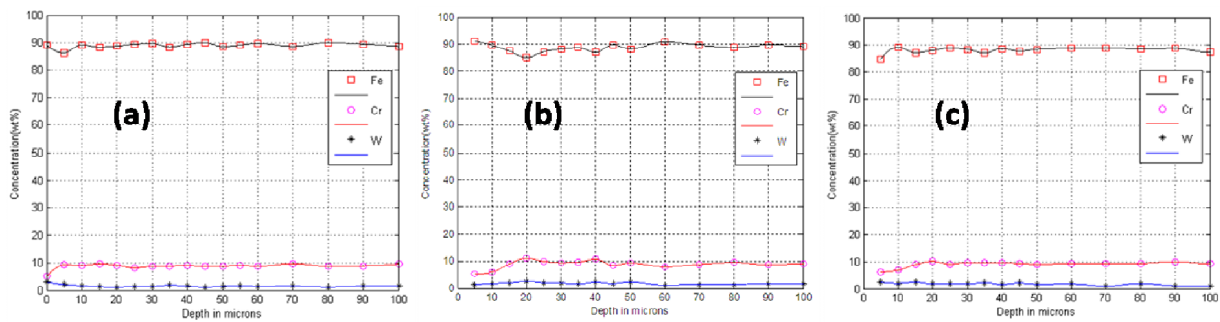

**Fig.4.** EDX line scan on cross section of IN-RAFMS after exposure to static Pb-Li at 550 C.
Exposure time: (a) 2500 hours, (b) 5000 hours and (c) 9000 hours.

## 3.5. Micro Vickers Hardness measurement

Micro Vickers hardness was measured on the cross section of the samples up to a depth of 100 μm and shown in figure 5 for all the three exposure times.

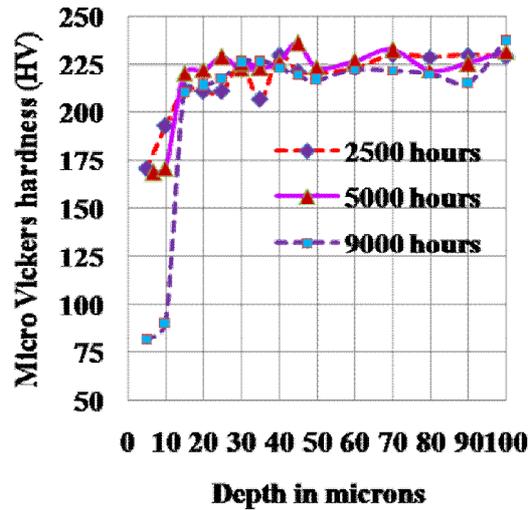

**Fig.5.** Depth profiles of Micro Vickers hardness of exposed IN-RAFMS samples to static Pb-Li at 550 C. Exposure time: (a) 2500 hours, (b) 5000 hours and (c) 9000 hours.

It is observed that hardness decreased in the near surface regions up to a depth of 15 μm in all the three cases. After 2500 and 5000 hours exposure to liquid metal, hardness decreased up to ~170 HV. After 9000 hours exposure, surface hardness decreased drastically up to 80 HV. Decrease in surface hardness was also observed for 9Cr-ODS steels exposed to static Li and Pb-Li at $600^o$ C for 250 hours by Y.Li et al.,[22] and for JLF-1 steel exposed to Li by Qi Xu et al.,[31].

## 3.6. Tensile testing

Tensile testing was carried out at room temperature on fresh and exposed samples using Instron make 5982 universal testing machine at a strain rate of $3 \times 10^{-3}$/sec. The results are tabulated in table 3. After exposure to Pb-Li, no significant change in the tensile strength was observed as compared to unexposed sample.

**Table 3: Tensile strength after exposure to Pb-Li**

| Sl. No | Exposure time (hours) | Tensile strength (MPa) |
|--------|-----------------------|------------------------|
| 1.     | Zero (unexposed)      | ~651                   |
| 2.     | 2500                  | 652.52                 |
| 3.     | 2500                  | 661.57                 |
| 4.     | 5000                  | 635.67                 |
| 5.     | 5000                  | 640.04                 |
| 6.     | 9000                  | 632.26                 |
| 7.     | 9000                  | 691.25                 |

## 4. Results and discussion

Dissolution of the alloying elements seems to be the main corrosion mechanism. From SEM/EDX measurements on the cross section of the samples, chromium was found selectively getting leached out from the near surface regions, ~15 microns from the surface. Chromium content reduced from 9% to (5-6)% after exposure to liquid metal. Tungsten was found high near the surface. This could be due to less solubility of tungsten in the liquid metal.

From the surface micrographs, pebble like structures were seen on the surface. When the surface of the structural material is in contact with Pb-Li eutectic, corrosion occurs on the surface and the bonding between the grains/sub grains becomes weak, leading to detachment of some grains/sub grains from the surface which pass into the eutectic.

From the Micro Vickers hardness measurements it was observed that the surface hardness decreased up to a depth of 15 μm for all the exposure times. Surface hardness decreased up to 170 HV after 2500 and 5000 hours exposure and up to 80 HV after 9000 hours exposure to liquid metal. From weight loss measurements, estimated corrosion rate was ~40 μm/year. Thus weight loss and decrease in surface hardness could be due to the dissolution of chromium in the liquid metal. Tensile strength of exposed samples to liquid metal remained almost same as that of fresh IN-RAFMS samples. It can be said that reduction of hardness in the near surface layers did not affect the tensile strength of the material.

**Acknowledgements**

Authors would like to acknowledge Mr. L.Narendrasinh L.Chauhan for supporting SEM measurements.